\documentclass[preprint,12pt]{revtex4}
\usepackage{epsfig}
\usepackage{natbib}
\usepackage[normalem]{ulem}  
\usepackage{xspace}
\usepackage{xcolor}
\usepackage{lineno}

 \usepackage{amsmath}
\usepackage{bm}

\begin{document}

\title{Interplay between scales in the nonlocal FKPP equation}
\begin{abstract}
\end{abstract}

\author{G. G. Piva}
\affiliation{Department of Physics, PUC-Rio, Rio de Janeiro, 22451-900 RJ, Brazil}
\author{E. H. Colombo}
\affiliation{Department of Ecology \& Evolutionary Biology, \\ Princeton University, Princeton, NJ 08544, USA\looseness=-1}
\affiliation{Department of Ecology, Evolution, and Natural Resources, \\ Rutgers University,  New Brunswick, NJ 08901, USA\looseness=-1}
\author{C. Anteneodo}
\affiliation{Department of Physics, PUC-Rio, Rio de Janeiro, 22451-900 RJ, Brazil}
\affiliation{INCT-Complex Systems, Brazil}

\begin{abstract}
We consider a generalization of the FKPP equation for the evolution of the spatial density of a single-species population where all the terms are nonlocal. That is,   the spatial extension of  each process (growth, competition and diffusion) is ruled by   an influence function, with a characteristic shape and range of action. Our purpose is to investigate the interference between these different components in pattern formation.  
We show that, while competition is the leading process behind   patterns, the other two can act either constructively or destructively. For instance, diffusion that is commonly known to smooth out the concentration field can actually favor pattern formation depending on the shape and range of the dispersal kernel.
The results are supported by analytical calculations accompanied by numerical simulations.
\end{abstract}
\maketitle



\section{Introduction}
\label{sec:intro}

Species spatial patterns are known to control several key ecological outcomes~\cite{tilman1997spatial}. These patterns can be induced by external constraints~\cite{hanski1999metapopulation} but more interestingly they can be a product of self-organization~\cite{camazine2020self,cross1993pattern}. Population self-organization has been shown, for instance, to improve the robustness and resilience of ecosystems~\cite{bonachela2015termite,Zhaoeabe1100}, and affect species coexistence~\cite{PhysRevLett.110.258101,Maciel21}. In biological populations,  the origins of  pattern formation has been typically linked to scale-dependent feedback mechanisms related to dispersal (diffusion), growth, and competition, among other processes~\cite{murray2001mathematical,rietkerk2008regular,cross1993pattern}. 
These elementary processes can be spatially-extended due to the dynamics of
other individuals and substances that act as mediators, 
 extending their effects across space. 
 In such a way, the reproduction, lifetime, and the  behavior of individuals depend on the surroundings. 
 For instance, this is the case of plants  when  water and wind mediate the growth and spread of the vegetation cover~\cite{hillerislambers2001vegetation,borgogno2009mathematical}. 
  
\begin{figure}[h!]
  \centering
  \vspace*{5mm}
   \includegraphics[width=0.9\columnwidth]{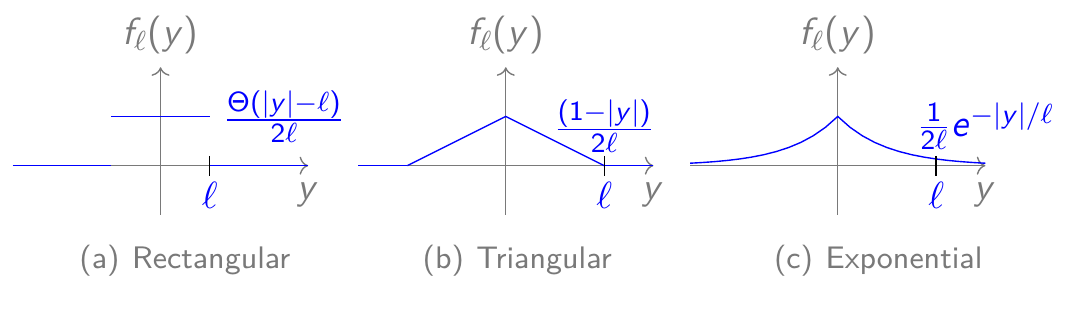}
  \caption{Typical kernels for the influence function $f_{\ell}(y)$ with interaction lengthscale $\ell$.}
  \label{fig:kernels}
\end{figure}

Our goal is to understand how the 
 different nonlocal processes interact and control self-organization.
For that purpose, we consider a single-species scenario, 
described by the nonlocal version of  the FKPP~\cite{FISHER1937WAVE} equation for the evolution of  the population density $u(x,t)$, whose original version includes standard diffusion, reproduction, and intra-specific competition with constant rates~\cite{murray2001mathematical}. We modify the FKPP equation by replacing the originally local terms by the convolution of the population density with influence functions
~\cite{murray2001mathematical,fuentes2003nonlocal,da2011pattern,colombo2012nonlinear}, obtaining
\begin{equation}
    \partial_t u(x,t) = D \overline{\nabla}^2 u + G[\mathcal{G}*u] -C u [\mathcal{C}*u] \,,
    \label{eq:fkpp_lin}
\end{equation}
where $D$,   $G$ and $C$ are positive coefficients, 
$\overline{\nabla}^2 u =[\mathcal{D}*u-u]$ is a generalized Laplacian that will be discussed below, and the influence functions, $\mathcal{D}(x)$, $\mathcal{G}(x)$ and $\mathcal{C}(x)$  embody the nonlocality of the diffusion, growth and competition processes, respectively, acting  through a convolution operation $f*u=\int_{-\infty}^{\infty} f(x-x') u(x',t) dx'$.
Each  influence function has a particular (normalized) shape, with a typical length-scale as main parameter, corresponding to an effective radius of action.
In Fig.~\ref{fig:kernels}, we  show typical kernels $f_\ell(y)$, characterized by a length-scale $\ell$, which can be used as  influence function of each process, $\mathcal {D}$, $\mathcal{G}$, $\mathcal{C}$, with respective scales $\ell=c$, $d$, and $g$. 
In all cases, the kernel is normalized, and, when $\ell\to 0$, it tends to a Dirac delta function.

Notice that, in Eq.~(\ref{eq:fkpp_lin}), 
the growth term, instead of being proportional to the local density, is proportional to the average of a neighborhood, resulting from the convolution of the density with the influence function  $\mathcal{G}(x)$, which accounts for facilitation~\cite{rietkerk2008regular}. Similarly, competition also occurs with a neighborhood (not necessarily the same as for growth regulation), weighted by  convolution with the function $\mathcal{C}(x)$. 
The operator $\overline{\nabla}^2 u =[\mathcal{D}*u-u]$ is a generalized Laplacian, governing nonlocal diffusion~\cite{kang2020age, bates2006some, andreu2010nonlocal}.  In experimental settings, the normalized kernel, $\cal{D}$, is associated with the probability distribution of the distance travelled by the walkers~\cite{turchin1998quantitative}. 
For a symmetric kernel of small width $d$, Taylor expanding $u(x,t)$ in the convolution integral leads to 
$\overline{\nabla}^2 u  \simeq  \frac{\mu_2}{2!} \partial^2_{x} u+ \frac{\mu_4}{4!} \partial^4_{x} u+\ldots$, where $\mu_n$ is the $n$th moment of the kernel.
Therefore, for small $d\ll 1$,  $\overline{\nabla}^2 u \simeq \mu_2\partial^2_{x} u$, which recovers the usual Laplacian diffusion with  coefficient proportional to the second moment of the dispersal kernel. 
 Furthermore, since $\mu_2 \sim d^2$, in this limit, the kernel width determines the rate of diffusion.

In the context of Eq.~(\ref{eq:fkpp_lin}), it is already known that nonlocal competition is key 
to pattern formation~\cite{fuentes2003nonlocal,martinez2013vegetation}.
In addition to competition, previous studies have also addressed the impact of  nonlocal facilitation and dispersal in pattern 
formation~\cite{borgogno2009mathematical}. However, it has typically been assumed that the  interactions governing facilitation and dispersal decay smoothly with distance (e.g., regulated by exponential and Gaussian kernels). 
In this case, the interplay between processes is rather trivial, with competition driving the formation of patterns and the others favoring the stability of the homogeneous state.
Here, we highlight that, for sufficiently compact kernels, there is a qualitative change in the conditions for spatial stability, which become determined by the (constructive or destructive) interference between the contributions associated to each process. This opens the possibility for a more complex scenario where the effects that a given process has on pattern formation cannot be known a priori without  information on the others involved. 
In the following sections, we investigate this interference phenomenon, focusing on the interplay between the interaction lengthscales associated to the elementary processes: $d$ (dispersal), $g$ (growth) and $c$ (competition), fixing different shapes for $\mathcal{D}$, $\mathcal{G}$ and $\mathcal{C}$ in Eq.~(\ref{eq:fkpp_lin}).

 The paper is structured as follows. In Section~\ref{sec:stability}, we perform the linear stability analysis of Eq.~(\ref{eq:fkpp_lin}). 
 The interference phenomenon is investigated in Section~\ref{sec:linear} for the simplest (rectangular) form of the influence function. 
A  nonlinear extension of Eq.~(\ref{eq:fkpp_lin}),   bringing new features to the effects of interference is analyzed in Sec.~\ref{sec:nonlinear}. Other kernels and variants are discussed in Sec.~\ref{sec:final}, containing final remarks.

\section{Linear stability analysis}
\label{sec:stability}

Considering a perturbation of the homogeneous solution of Eq.~(\ref{eq:fkpp_lin}), of the form $u(x,t)=u_0+\epsilon(x,t)$, with $u_0=G/C$ and 
 $ \epsilon(x,t)= a \exp(i k x+\lambda t)$, where $a<<u_0$, and substituting it into Eq.~(\ref{eq:fkpp_lin}), we have  
 \begin{align}
    \nonumber \partial_t \epsilon=D  [\mathcal{D}*\epsilon-\epsilon] + G [\mathcal{G}*\epsilon -    \mathcal{C}*\epsilon-\epsilon] = \lambda(k) \,\epsilon. 
    \label{eq:fkpp_lin}
\end{align}
Therefore, the growth rate of mode $k$ is given by
\begin{equation}
    \lambda(k)=
    \underbrace{D(\tilde{\mathcal{D}}-1)}_{\textstyle \lambda_d}
     +\underbrace{G \tilde{\mathcal{G}}}_{\textstyle \lambda_g}
  +\underbrace{(-G) (\tilde{\mathcal{C}}+1 )}_{\textstyle \lambda_c},
    \label{eq:dispersion}
\end{equation}
where $\tilde{f}(k) = \int_{-\infty}^{+\infty} f(x) e^{-i kx} dx$ is the Fourier transform of the influence function $f$.
For locality, when $f(x)$ is a Dirac delta, we have $\tilde{f}(k)=1$.

 Note that if $\lambda(k)<0$, for all $k$, then the homogeneous state is stable. Otherwise, we have spatial instability. 
 For the emergence of    regular  spatial structures, a typical expected signature is that there exists a  range of modes for which $\lambda(k)>0$, with a clear dominant mode  $k^*$~\cite{cross1993pattern},  together with  $\lambda(0)<0$ to stabilize the reference level.  
 Furthermore, the mode $k^*$, which is the initially fastest growing one, is frequently the mode  that dominates in the long-time non-homogeneous state~\cite{colombo2012nonlinear}.  This indicates weak mode interaction and allows to use the short-time dynamics as a proxy for the longtime patterns.

Assuming that the influence functions for the three processes decay with distance and are normalized, Eq.~(\ref{eq:dispersion})   leads to spatial patterns only when competitive interactions are sufficiently  retricted within a certain radius of influence, $c$.
For instance, assuming that solely competition is nonlocal in Eq.~(\ref{eq:fkpp_lin}) (i.e., with local growth and dispersal), and proposing the choice  $\mathcal{C}(x) \sim \exp{(-|x|^{\alpha})}$, it has been shown that the population can self-organize into periodic structures only if $\alpha>2$, which corresponds to platykurtic functions.
Similarly, for $\mathcal{C}$ belonging to the $q$-exponential family, it has been shown that pattern formation is possible only for sub-triangular kernels~\cite{Dornelas2021}. 
This requirement of compactness  is associated to its  manifestation  in Fourier space. The crucial point is that sufficiently compact kernels  have non-positive definite  Fourier  transform, allowing changes of  sign in  $\lambda(k)$~\cite{PhysRevLett.98.258101}. For instance, while the exponential kernel has  Fourier transform  $\tilde{f}(k) \propto  1/(1 + (k\ell)^2)$ which is positive, and monotonic for $k>0$, the most compact form, the rectangular influence function, has Fourier transform $\tilde{f}(k)=\sin(k\ell)/(k\ell)$, which takes positive and negative values, with decaying oscillations.

Although dispersal and growth are not responsible for pattern formation,   if they introduce oscillatory behavior in  $\lambda(k)$,   there might be  interference between the different oscillating components $\lambda_d$, $\lambda_g$ and $\lambda_c$.

\section{Interference and its consequences}
\label{sec:linear}

In the subsequent analytical and numerical studies, we consider only influence functions of rectangular shape, like the one shown in Fig.~\ref{fig:kernels}(a). As mentioned previously, this choice is a representative case of the effects of ``compact" kernels on the spatial stability. 
Other shapes are discussed in Sec.~\ref{sec:final}.

\begin{figure}[h!]
  \vspace*{5mm}
 \centering
    \includegraphics[width=0.48\columnwidth]{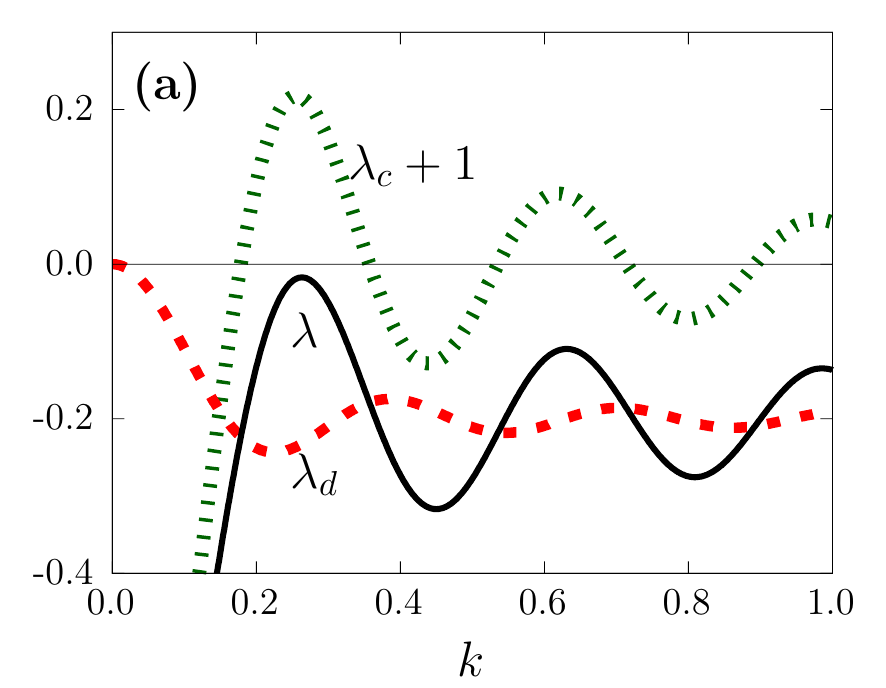}  \includegraphics[width=0.48\columnwidth]{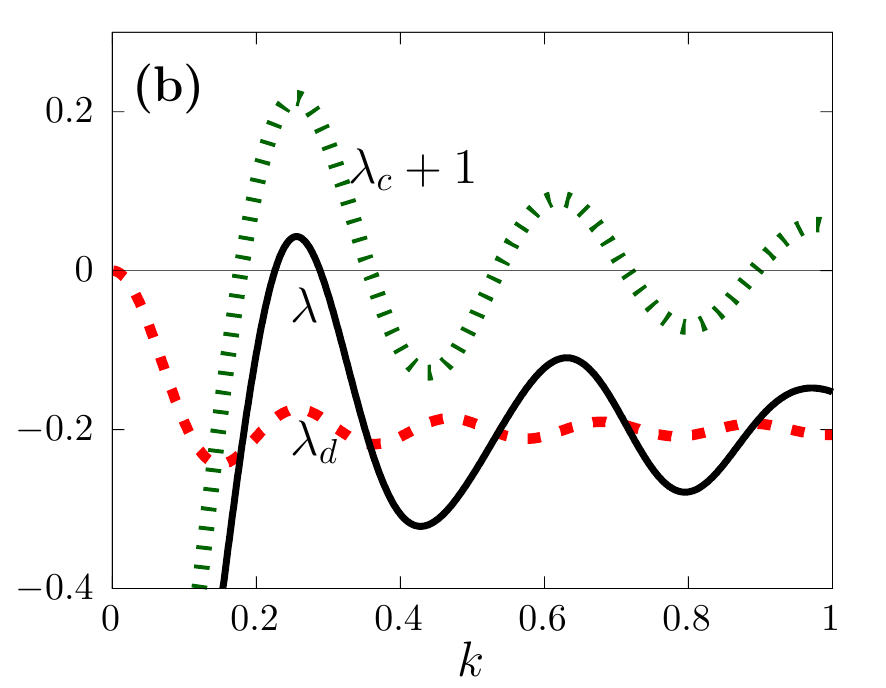}
\caption{Interference analysis according to the decomposition of Eq.~(\ref{eq:dispersion}), for rectangular $\cal{C}$ and $\cal{D}$, while the growth is local ($\tilde{\mathcal{G}}(k)\to 1$), hence $\lambda(k)=\lambda_d + \lambda_c+1$ and we plot $\lambda$, $\lambda_d$ 
and $\lambda_c+1$. 
Parameters are  $d=20$ (a),   $d=30$ (b), 
and  fixed $c=18$  and $D=0.2$. 
Notice that the absolute maximum of  $\lambda(k)$,  changes from negative (a) to positive (b)  when the scale $d$ varies, due to   destructive and constructive combinations, respectively. 
}
    \label{fig:interference}
\end{figure}

Using the fact that the Fourier transform of the rectangular kernel is the  function ${\rm sinc}(x)=\sin(x)/x$, then Eq.~(\ref{eq:dispersion}) becomes
\begin{equation}
    \lambda(k)=
    D\bigl(  {\rm  sinc} (k d)-1 \bigr)
    +
   G\; {\rm sinc}(kg)   
   -
   G \bigl(  {\rm sinc}(kc) +1 \bigr).
   \label{eq:dispersionSQ}
\end{equation}
 Let us first assume $g\to0$ (hence, ${\rm  sinc}(kg) \to 1$), implying local growth, and focus on the interplay between the competition and diffusion scales. 
In Fig.~\ref{fig:interference}, we plot $\lambda$ (solid line) and its components $\lambda_d$ and 
$\lambda_c$  vs. $k$, for two values of $d$.

\begin{figure}[h!]
   \centering
    \includegraphics[width=0.48\columnwidth]{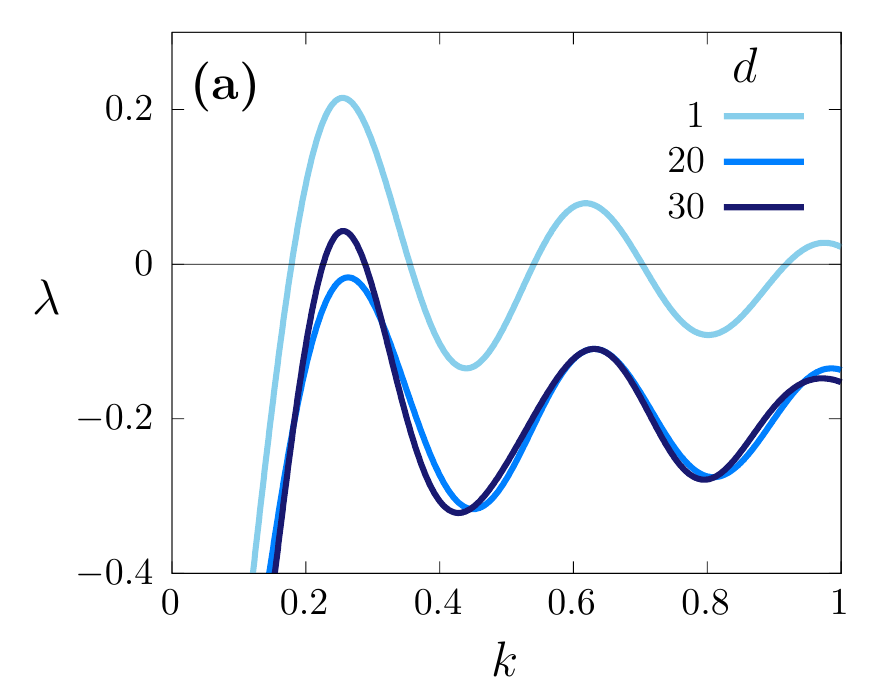}  
  \includegraphics[width=0.48\columnwidth]{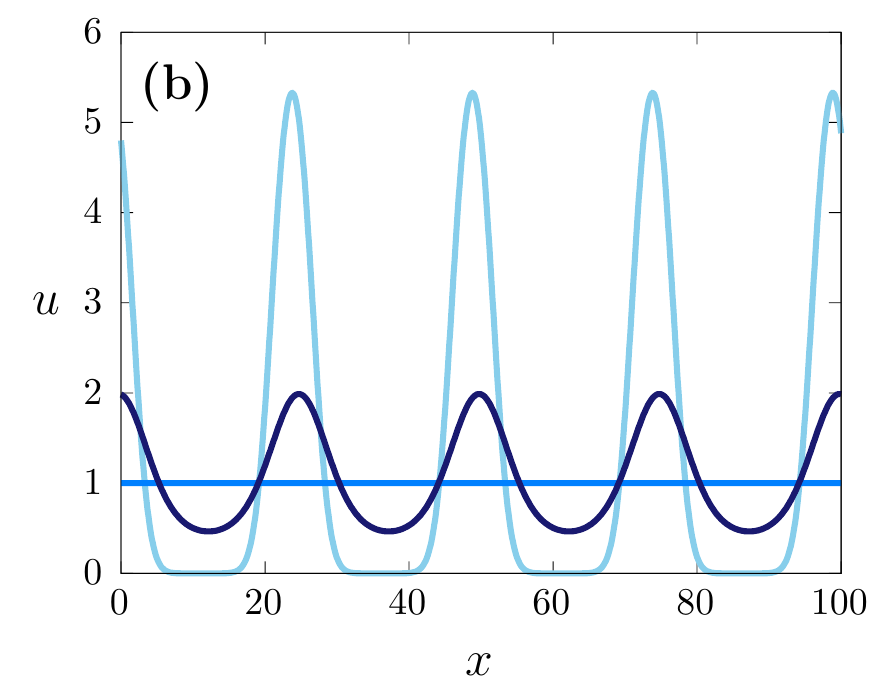}
    \caption{(a)  Mode growth rate $\lambda(k)$ and (b) long-time density profile $u(x)$, for different values of $d$ indicated in the legend (the darker the larger), with  $c=18$ (diffusion and competition kernels are rectangular, while growth is local, i.e., $g\to 0$), with coefficients $D=0.2$ and $G=1$.  
    The profiles are the results of simulations for a box of $L=100$ with periodic boundary conditions, at $t=100$.
    }
    \label{fig:lambda_density_D}
\end{figure}

In Fig.~\ref{fig:interference}(a), we show an example where the oscillations introduced by competition and diffusion are out of phase   around the global maximum at $k^*\simeq 0.25$,  hence, there is  destructive interference lowering the peak, and even yielding  $\lambda(k^*)<0$.
In contrast, in Fig.~\ref{fig:interference}(b), the components interfere constructively,  rising the peak, and  yielding $\lambda(k^*)>0$.  
This example shows how diffusion can favor the emergence of patterns, depending on its range of action $d$. This contrasts with the fact that diffusion becomes  deleterious for increasing intensity $D$, which lowers the level around which $\lambda_d$ oscillates, and hence sinks the dominant peak.

Actually, $d$ has a non-monotonic effect on pattern formation, which is illustrated  in Fig.~\ref{fig:lambda_density_D}, where  we plot $\lambda(k)$ in panel (a), and  the corresponding long-time profiles  of $u(x)$ in panel (b). These profiles were  
obtained from the numerical integration of Eq.~(\ref{eq:fkpp_lin}), in a box of size $L=100$ with periodic boundary conditions~\cite{note:numerical}, departing from a small random perturbation around the homogeneous state. 
Notice that, for a small value of $d$ close to locality ($d=1$ in the figure), there is spatial instability  of the homogeneous state, i.e., $\lambda(k^*)>0$, producing a well defined pattern that can be observed in Fig.~\ref{fig:lambda_density_D}(b)  (lighter blue). As $d$ increases, the interference becomes   destructive until making $\lambda(k^*)<0$,  but further increasing $d$ there is again constructive interference producing $\lambda(k^*)>0$ (darker blue).  The consequence of this alternation in the positiveness of the absolute maximum is reflected in the long-time profiles in Fig.~\ref{fig:lambda_density_D}(b). Also note that the dominant wavelength of the long-time patterns, $\Lambda \simeq 25$, is in good agreement with the value $2\pi/k^*$ (with $k^*\simeq 0.25$) predicted by  the linear stability analysis.

\begin{figure}[b!]
   \centering
    \includegraphics[width=0.5\columnwidth]{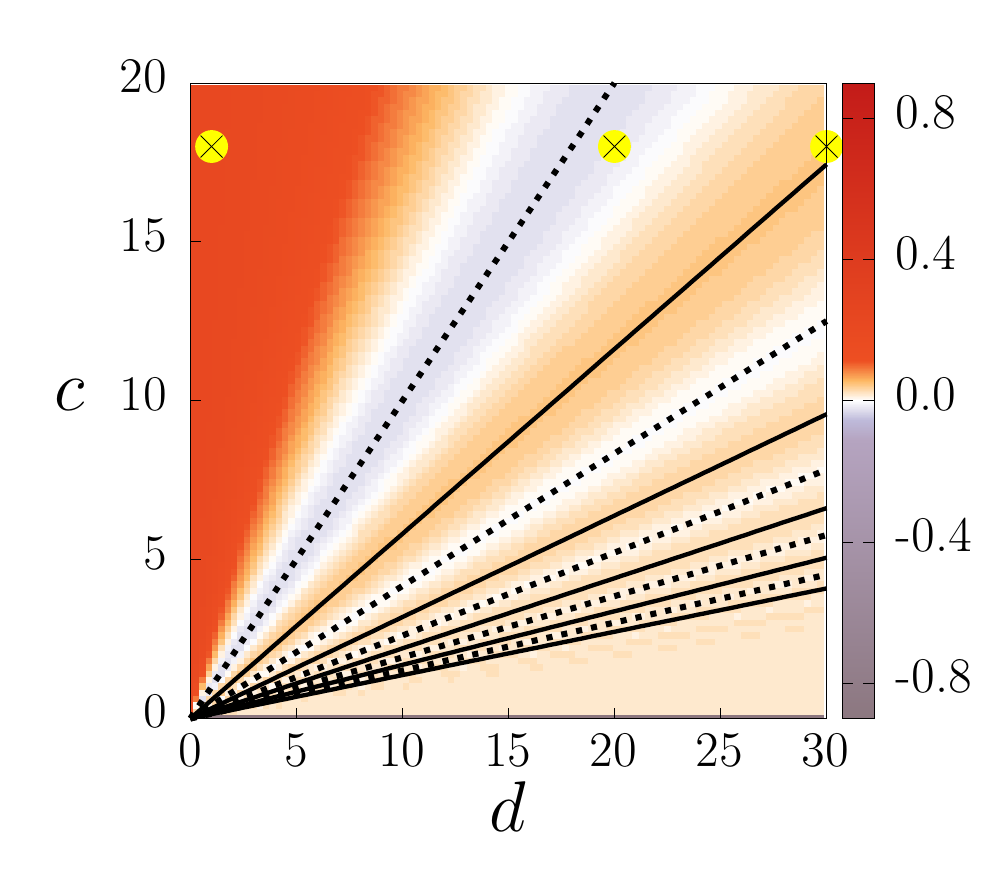}  
\caption{Heat plot of the maximal growth rate $\lambda(k^*)$, for 
rectangular competition and diffusion kernels, and local growth ($g\to 0$), with $G=1$, $D=0.2$.    
The straight lines correspond to local maxima (solid) and minimal (dashed), predicted from the extrema of the sinc function as explained in the text. The symbols $\times$ correspond to the cases shown in Fig.~\ref{fig:lambda_density_D}. 
}
    \label{fig:linhas}
\end{figure}
 
In order to identify the set of scales $(c,d)$ for 
which constructive and destructive interference occur, it is important to first note that the contribution of competition $\lambda_c$ has a maximum at $k\simeq 1.43 \pi/c$,
which is the first minimum of the sinc function ($k\simeq 0.255$ in the case of 
Fig.~\ref{fig:interference}). This is the contribution that determines the wavelength of the patterns that emerge if the peak were positive.
Then, note that the component $\lambda_d$ acts constructively,
 when one of its maxima  
(which are related to the maxima $x_M$ of ${\rm sinc}(x)$,   with 
$x_M/\pi = 2.46, 4.47, 6.47...\simeq 2n+1/2$, for $ n=1,2,3,\ldots$),
coincide with the first peak of $\lambda_c$. 
This occurs for $d = x_M/(1.43\pi) c$.
Analogously, the contribution of $\lambda_d$ is destructive when  
 $d = x_m/(1.43\pi) c$, where the minima $x_m$ of 
 ${\rm sinc}(x)$ are given by $x_m/\pi = 1.43, 3.47,..., \simeq 2n-1/2$, for $ n=1,2,3,\ldots$. 
These results explain the interspersed regions of  instability (orange) and stability of flat profiles (lilac) in the plane of parameters $d-c$ seen in Fig.~\ref{fig:linhas}.

\begin{figure}[h!]
   \centering
    \includegraphics[width=0.33\columnwidth]{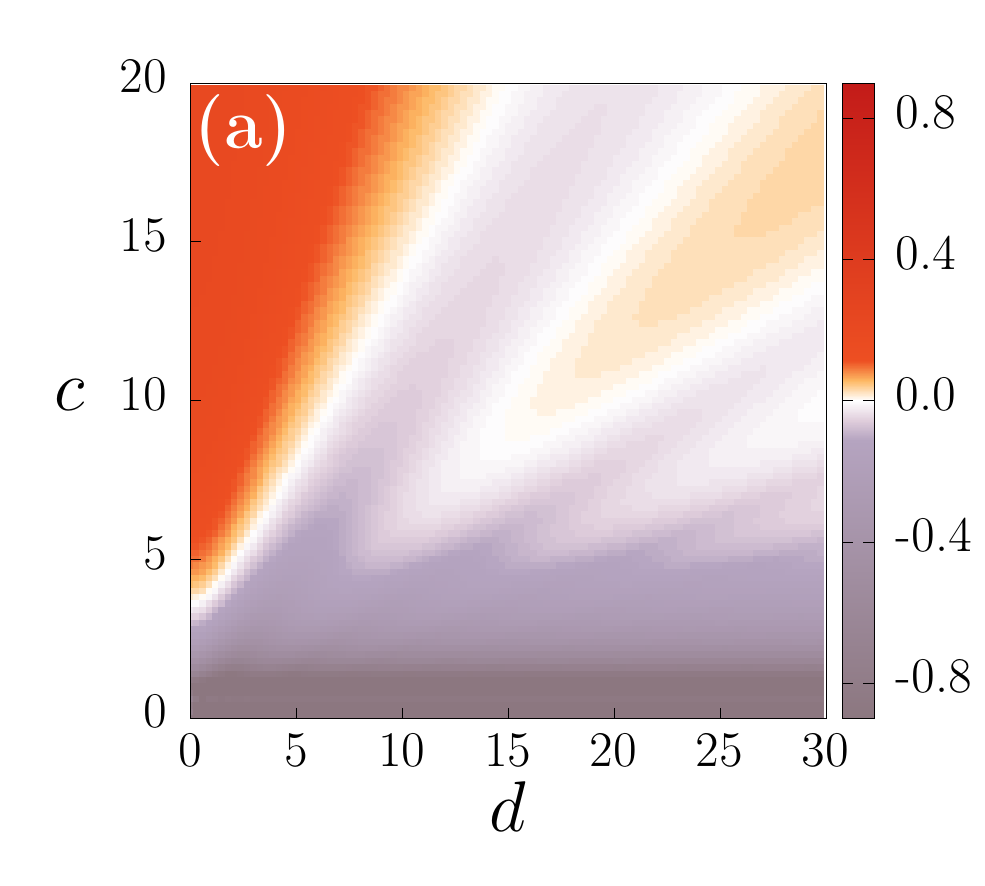}\includegraphics[width=0.33\columnwidth]{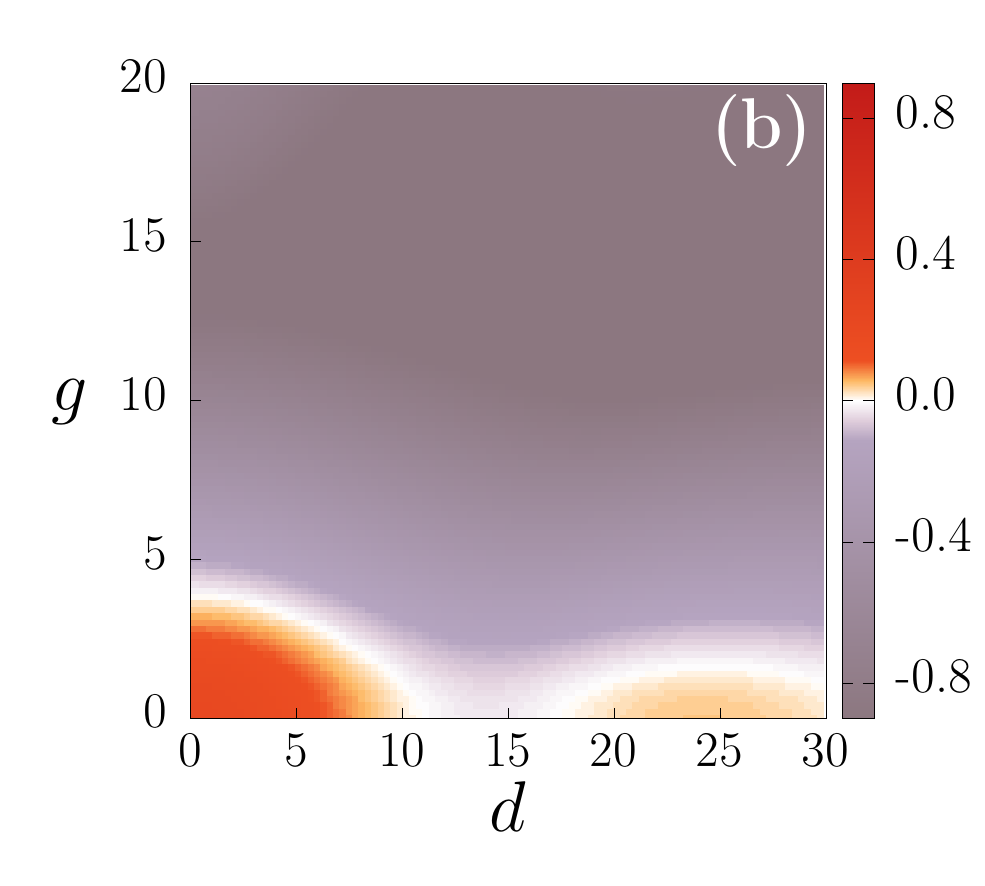}\includegraphics[width=0.33\columnwidth]{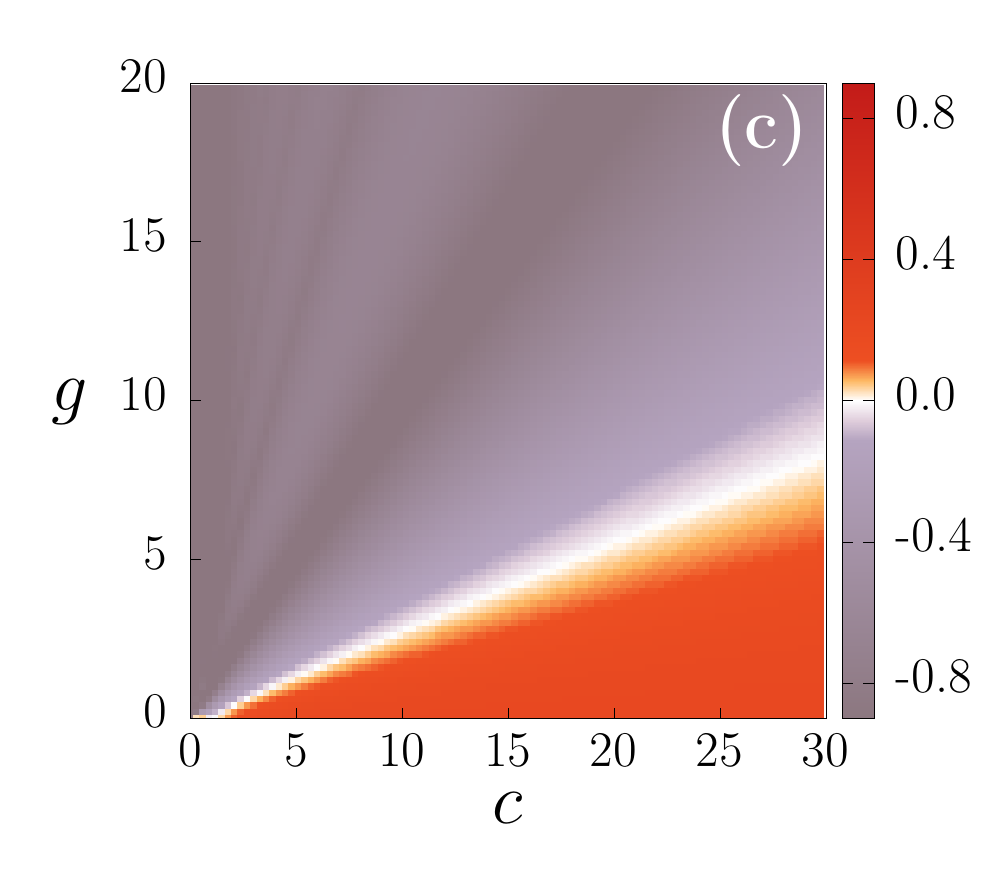} 
    \caption{Heat plots of the maximal mode-growth rate $\lambda(k^*)$ in different planes of the space of the three (rectangular) kernel scales  (a) $d-c$ plane, at $g=1$, (b) $g-d$, at $c=14$ (c) $g-c$ plane  plane, at $d=1$. Growth and diffusion coefficients are $G=1$ and $D=0.2$, respectively.}
    \label{fig:diagrams_SQ}
\end{figure}

A broader portrait, considering the fully nonlocal model, varying all the scales ($d,g$ and $c$) in the range (0,30], is shown in 
Fig.~\ref{fig:diagrams_SQ}.  Each panel contains heat plots of 
the maximal mode-growth rate $\lambda(k^*)$ in different planes of the $\{d,g,c\}$ parameter space. 
In panel (a), we show the effect of short-range facilitation, adding a pinch of nonlocality to growth, in contrast to the growth-local Fig.~\ref{fig:linhas}.
The straight lines
observed in Fig.~\ref{fig:linhas} are distorted in Fig.~\ref{fig:diagrams_SQ}(a)  
due to the fact that $\lambda_g$ is now mode-dependent.
The non-monotonic behavior of $\lambda(k^*)$ is preserved when $g>0$, 
as can be observed when $c$ or $d$ vary along cuts of the diagrams. 
Curiously, this non-monotonicity is not observed when $g$ varies, presenting a single transition between pattern to no-pattern regions.  
However, notice the variations of intensity in the   lilac  regions (without patterns), suggesting that the hidden structure also reflects the interference of scales.

\begin{figure}[h!]
 \centering
    \includegraphics[width=0.5\columnwidth]{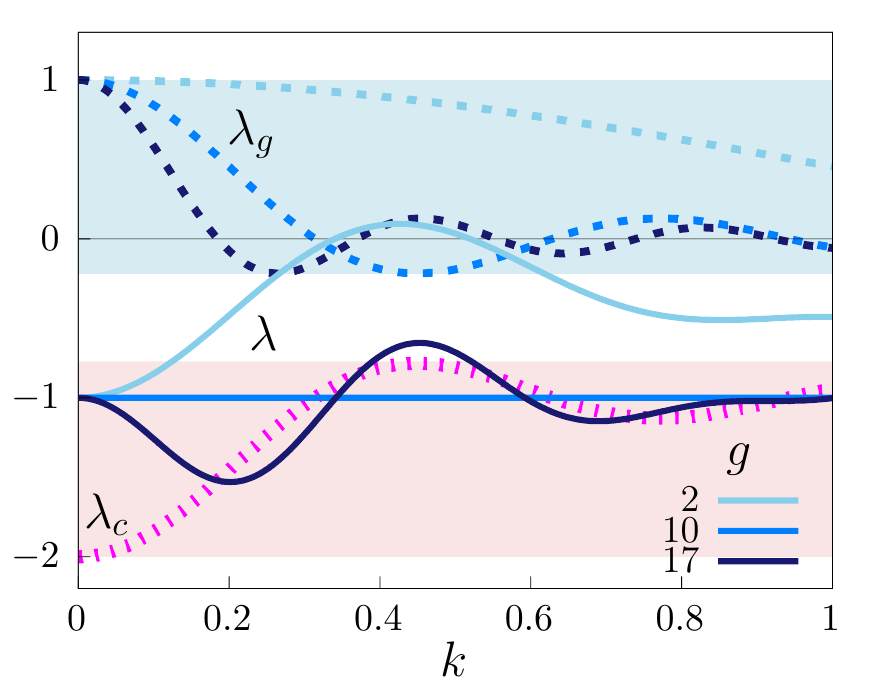}
\caption{Interference analysis for different scales $g$ indicated in the legend, with  fixed scale $c=10$ (for growth and competition rectangular kernels), $D=0$ and $G=1$.  $\lambda=\lambda_g+\lambda_c$ (solid lines) and components  $\lambda_g$ (dashed lines) and  $\lambda_c$ (dotted line) vs. $k$.  The shadowed regions highlight the range of variation of the components $\lambda_g$ and $\lambda_c$. 
}
    \label{fig:interferenceG}
\end{figure}

Finally, the effect of varying $g$ is illustrated in 
Fig.~\ref{fig:interferenceG}. 
Notice that for  $g \ll c$, $\lambda_g$ contributes to pattern formation  since it adds $\lambda_g\simeq 1 $ to the peak of $\lambda_c$.  
However,  increasing $g$ shrinks the profile of $\lambda_g$, tending to lower the value of $\lambda(k^*)$ below zero, hence spoiling patterns (lilac regions in Fig.~\ref{fig:diagrams_SQ}).  
Fig.~\ref{fig:interferenceG} remarks that the facilitation component  $\lambda_g$ combined with $\lambda_c$ also produces an interference phenomenon. However,  growth can not produce the alternate behavior seen for diffusion. 
This is why the phenomenon was not observed in Ref.~\cite{da2011pattern}, where only nonlocal growth and nonlocal competition were considered.

In fact, the distinct effects of growth and dispersal 
come from the bounds of each contribution (highlighted by shadowed bands in Fig.~\ref{fig:interferenceG}). 
The growth component $\lambda_g$ is predominantly positive, oscillating  in  the interval 
$(-0.217G,G]$ (see Fig.~\ref{fig:interferenceG}), 
hence it favors a positive $\lambda(k^*)$, only if $g$ is small enough in comparison to $c$ (case $g=2$ in the figure), otherwise, the resulting $\lambda$ is negative. 
Differently, the dispersal component, 
$\lambda_d=D({\rm sinc}(kd)-1)$, is always non-positive, oscillating within the interval $(-1.217D,0]$.

\section{Revealing hidden interference}
\label{sec:nonlinear}

In the previous section, we have investigated  the interference of the kernels. While dispersal presented a significant effect giving rise to the interspersed regions seen in Fig.~\ref{fig:linhas} in the plane of lengthscales $d-c$, the  role of facilitation, regulated by $g$, was not so rich. 
However, as seen in Fig.~\ref{fig:interferenceG}, it is clear that, for the growth process, 
the interference mechanism is also present but hidden. 
 It is hidden in the sense that the height of the main peak of $\lambda(k)$ changes non-monotonically with $g$, 
 but it is always negative.  
This opens the question about whether some internal or external contribution to $\lambda(k)$ could bring the lilac darker regions in Fig.~\ref{fig:diagrams_SQ}(c) to the surface ($\lambda(k^*)>0$).
This kind of effect has been observed when certain types of noise are present~\cite{da2014effect}.  In the following extension of our model we explore that nonlinearities can have similar impact.

We now generalize the nonlocal model assuming that the neighborhood acts nonlinearly, i.e., that the rates of the processes are not proportional to the density averaged over the influence function, but to its power, with  an arbitrary  exponent. 
In fact, in many real problems, the coefficients are density-dependent, representing internal feedbacks~\cite{martinez2008continuous,martinez2009generalized,cabella2011data,dos2014generalized,dos2015models,colombo2018nonlinear}, or the heterogeneity of the environment~\cite{kath1984waiting}.

Taking these considerations into account, we write the evolution equation
\begin{equation}
    \partial_t u(x,t) = D \overline{\nabla}^2 u^\nu + G[\mathcal{G}*u]^\gamma -C u^\alpha [\mathcal{C}*u]^\beta \,,
    \label{eq:fkpp_nonl}
\end{equation}
where $\overline{\nabla}^2 u^\nu = [\mathcal{D}*u^\nu-u^\nu]/\nu$, with $\nu>-1$, is the generalized Laplacian of $u^\nu(x,t)$, that is the nonlocal generalization of nonlinear diffusion~\cite{colombo2012nonlinear}, and the exponents 
$\alpha$, $\beta$ and $\gamma$ are positive exponents, verifying $\gamma < \alpha+\beta$ to forbid unlimited growth.

Substituting the perturbation of the homogeneous state, $u(x,t)=u_0+\epsilon(x,t)$, into Eq.~(\ref{eq:fkpp_nonl}), where $u_0=\left(G/C\right)^{1/(\alpha+\beta-\gamma)}$, and neglecting terms of order higher than one in  $\epsilon(x,t)$, we obtain
\begin{equation}
    \lambda(k)=D u_0^{\nu-1} (\Tilde{ \mathcal{D}}-1)+G\gamma u_0^{\gamma-1} \Tilde{\mathcal{G}}-C u_0^{\alpha+\beta-1} (\beta \Tilde{\mathcal{C}}+\alpha ).
    \label{eq:general_dispersion}
\end{equation}
In the nonlinear case, the level $u_0$ may lead to different results. In particular, for  $u_0=1$, Eq.~(\ref{eq:general_dispersion}) becomes
\begin{equation}
    \lambda(k)=
    \underbrace{D  (\Tilde{ \mathcal{D}}-1)}_{\lambda_d}+
    \underbrace{G\gamma  \Tilde{\mathcal{G}}}_{\lambda_g}
    +\underbrace{(-G)(\beta \Tilde{\mathcal{C}}+\alpha )}_{\lambda_g}. 
    \label{eq:general_dispersion1}
\end{equation}

\begin{figure}[h!]
 \centering
    \includegraphics[width=0.45\columnwidth]{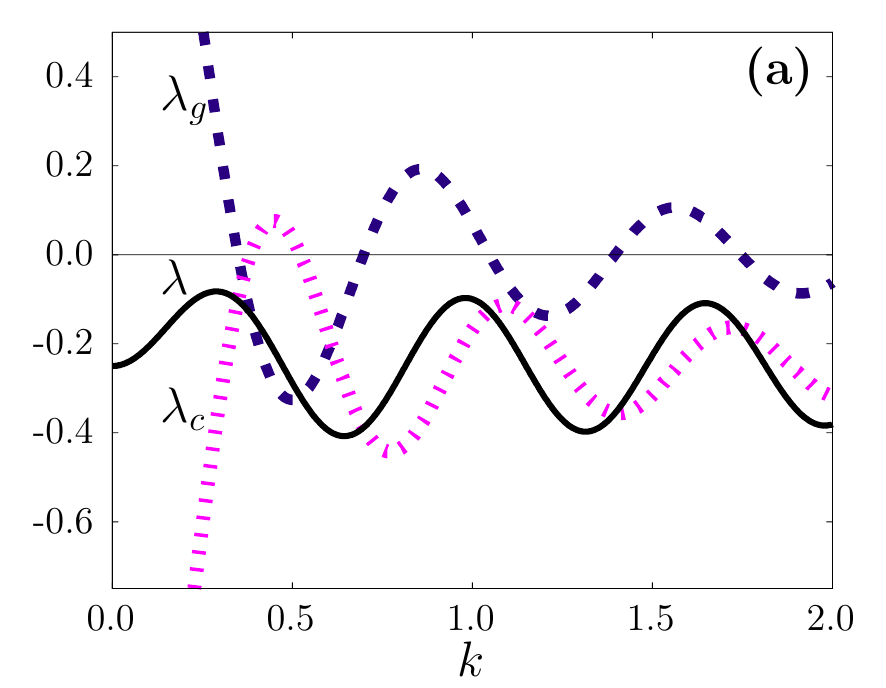} \includegraphics[width=0.45\columnwidth]{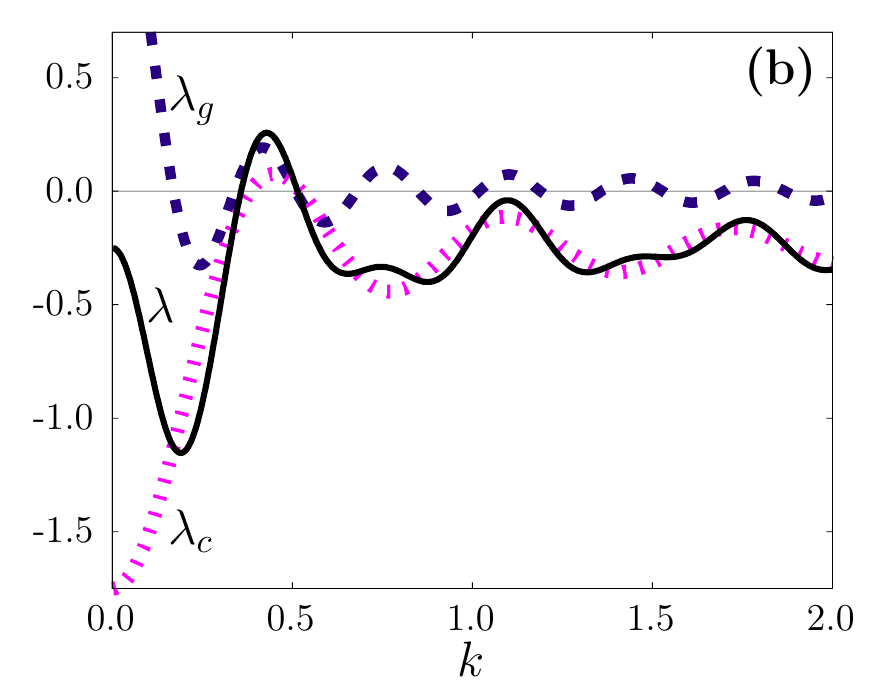}
\caption{Interference analysis for rectangular $\cal C$ and $\cal D$ with (a) $g= 10$ (b) $g= 18.5$, and $c=10$, with coefficient $G= 1$, and negligible diffusion with $D=0$.  Notice that the absolute maximum of $\lambda(k)$, changes from negative (a) to positive (b) when the scale $g$ varies, due to the possible destructive and constructive roles, respectively. Moreover, $u_0=1$, and the exponents are $\alpha=0.25$ and $\beta=\gamma=1.5$.}
    \label{fig:interference_nonlinear}
\end{figure}

 Eq.~(\ref{eq:general_dispersion1}) shows that  the exponents that characterize the nonlinearities  appear controlling  the strength of the components $\lambda_g$ and $\lambda_d$. 
This permits that the oscillations of the growth component $\lambda_g$ interfere in $\lambda(k)$ differently to the linear case where growth and competition have the same intensity $G$  Eq.~(\ref{eq:dispersion}). 
The distinct role of $\lambda_g$ in the nonlinear case is shown in Fig.~\ref{fig:interference_nonlinear}, where the diffusive term has been neglected.
In  Fig.~\ref{fig:interference_nonlinear}(a), for $g=10$, 
there is destructive interference and $\lambda(k^*)$ is negative, while 
in Fig. \ref{fig:interference_nonlinear}(b), for $g=18.5$, 
the growth component contributes constructively and rises the peak $\lambda(k^*)$ above the zero baseline.

\begin{figure}[h!]
   \centering
    \includegraphics[width=0.45\columnwidth]{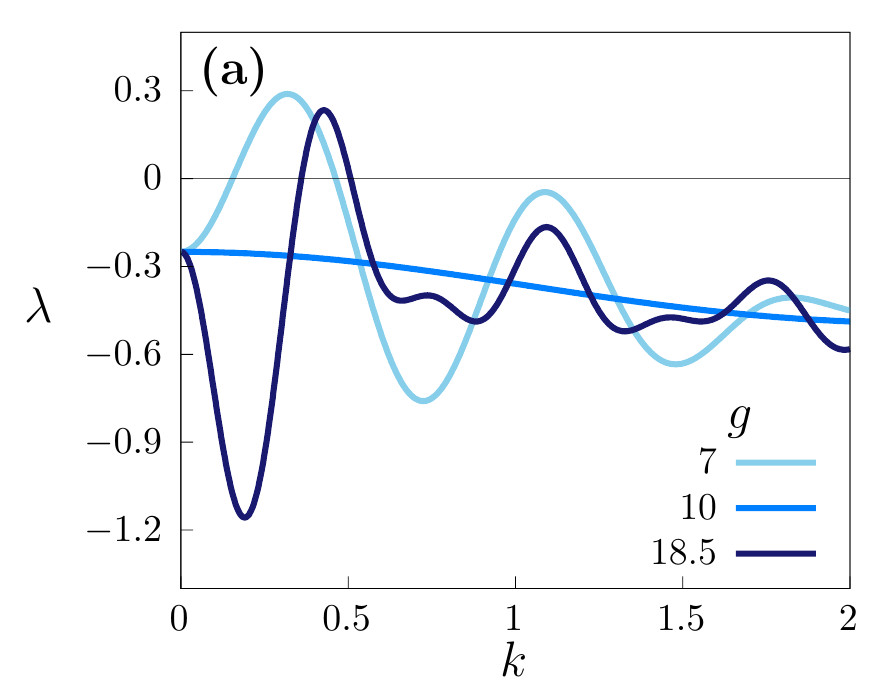} \includegraphics[width=0.45\columnwidth]{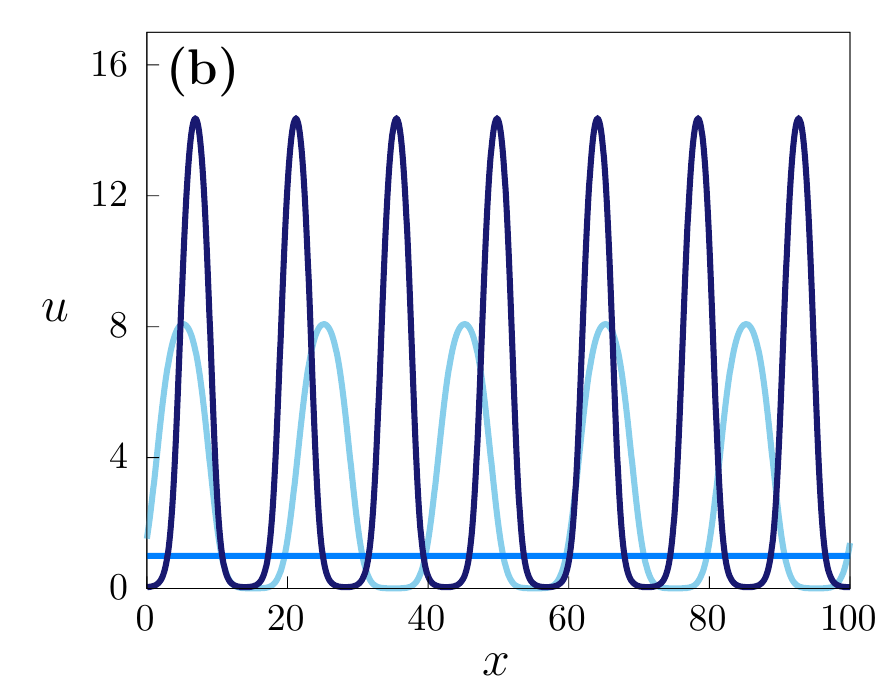}
    \caption{(a) Growth rate $\lambda(k)$ and  (b) long-time density profile $u(x)$, for different values of $g$ indicated in the legend (the darker the higher). 
    The other parameters are $c=10$, $d=2$ (all rectangular kernels), with coefficients $G=1$ and $D=0.2$.  
    The profiles are the results of simulations for a box of $L=100$ with periodic boundary conditions, at $t=200$. Exponents: $\alpha=0.25$, $\nu=1$ and $\beta=\gamma=1.5$
    Moreover, $u_0=1$, and the exponents are $\nu=1$, $\alpha=0.25$ and $\beta=\gamma=1.5$
    }
    \label{fig:nonlinear_density}
\end{figure}
\begin{figure}[h!]
    \includegraphics[width=0.33\columnwidth]{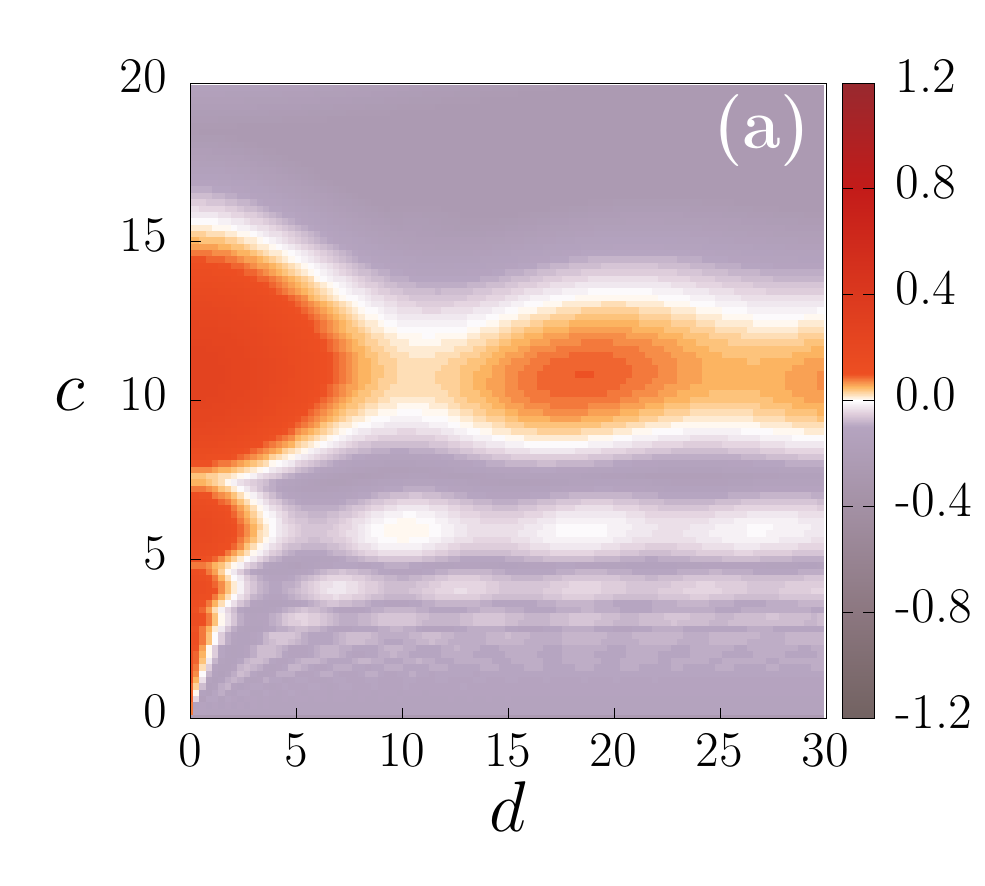} \includegraphics[width=0.33\columnwidth]{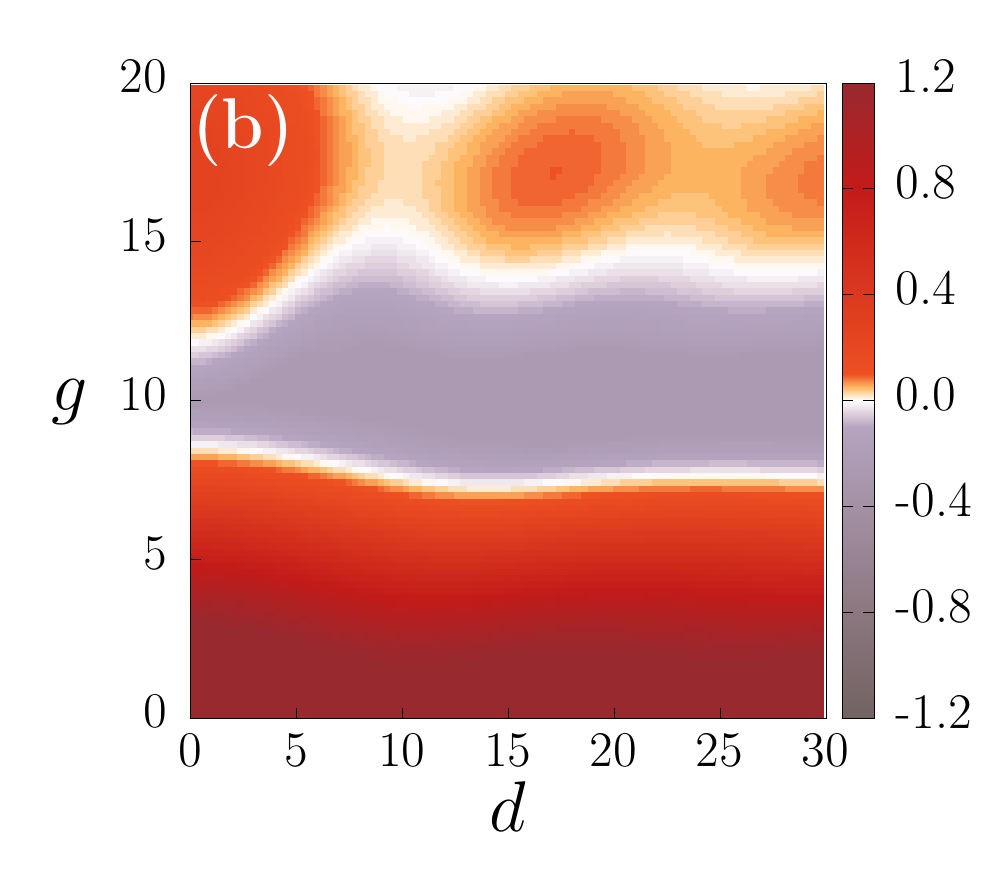}\includegraphics[width=0.33\columnwidth]{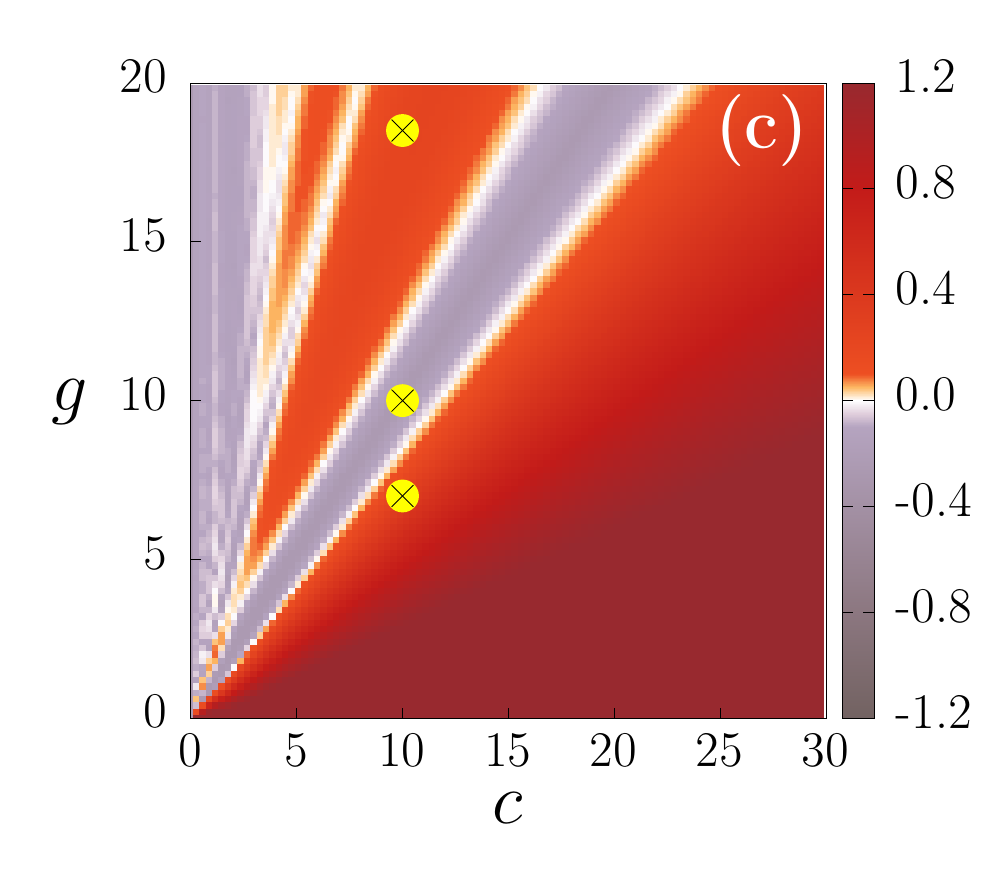} 
    
    \caption{Maximal growth rate  $\lambda(k^*)$ for rectangular kernels in the planes   (a) $d-c$, with $g=18.5$,  (b) $d-g$, with $c=10$, (c) $c-g$, with $d=2$. Coefficients are $D=0.2$ and $G=1$, $u_0=1$, and exponents are $\nu=1$, $\alpha=0.25$, $\beta=\gamma=1.5$. The symbols $\times$ in (c) correspond to the cases shown in Fig.~\ref{fig:nonlinear_density}. 
    }
    \label{fig:diagram_nonlinear}
\end{figure}

In Fig.~\ref{fig:nonlinear_density}, we show  the longtime profiles for increasing facilitation range. Due to the presence of interference, we see an alternation between patterned and homogeneous longtime distributions.
A broader picture of the phase diagram in the space of the interaction scales are shown in the heat plots of Fig.~\ref{fig:diagram_nonlinear}.

\begin{figure}[h!]
        \includegraphics[width=0.45\columnwidth]{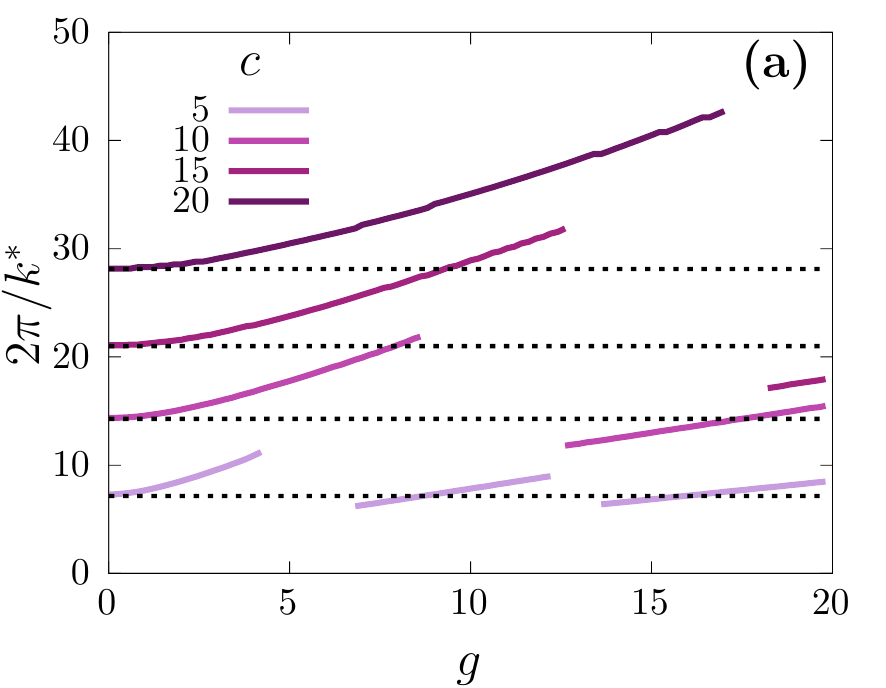} \includegraphics[width=0.45\columnwidth]{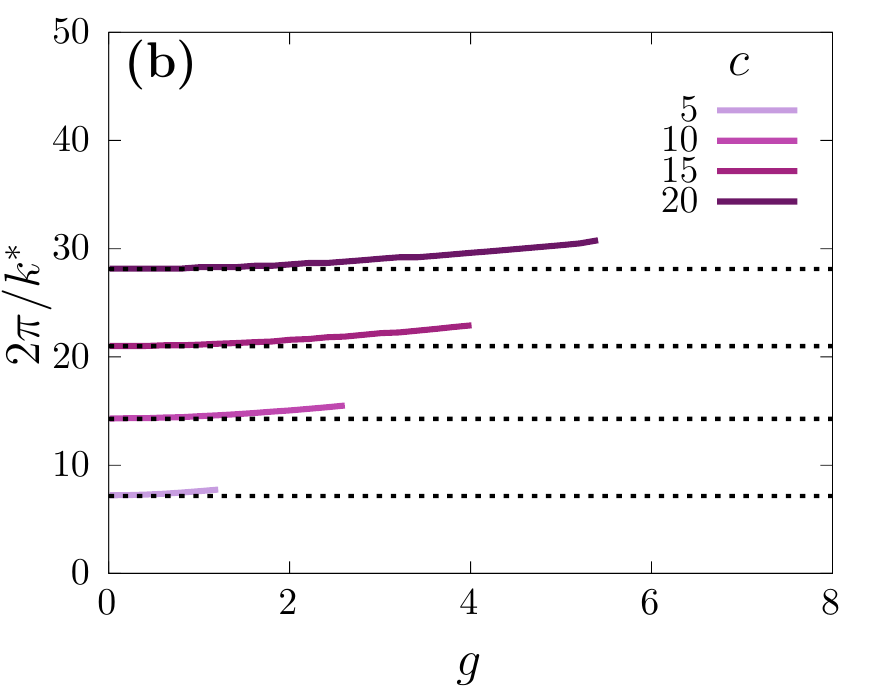}    
       \caption
    {Dominant pattern wavelength $2\pi/k^*$. 
    (a) Nonlinear case: corresponding to cuts in Fig.~\ref{fig:diagram_nonlinear}(c) at several values of $c$ indicated in the legend. 
    (b) Linear case:  corresponding to cuts of  Fig.~\ref{fig:diagrams_SQ}(c) at the same values of $c$. 
    The dotted lines correspond to the prediction in the local limit. 
    While in the linear case (b) the wavelength depends essentially only on $c$, in the nonlinear case (a) can vary significantly with other scales.  
    }
    \label{fig:wavelength}
\end{figure}

Another effect arising from the nonlinearity introduced in this section is that the wavelength of the resulting patterns can vary significantly with scales other than that determined by the competition kernel. 
This is illustrated in Fig.~\ref{fig:wavelength}, where we contrast the effect of facilitation spatial scale, $g$.
In Fig.~\ref{fig:wavelength}(a), we exhibit a significant variation of the wavelength  $\Lambda=2\pi/k^*$ as a function of  $g$, for fixed values of $c$, with the same values of the parameters used in Fig.~\ref{fig:diagram_nonlinear}(c). 
These outcomes are in contrast with those produced by the linear version of the model, as exemplified in Fig.~\ref{fig:wavelength}(b).  
In the latter case,  the wavelengths have small deviations from the value determined by the first peak in $\lambda_c$, that 
is $\Lambda=c/0.715 $ (horizontal dotted lines). 
Meanwhile in the former (nonlinear) case, the deviations are significant, which is possible due to the role of facilitation.   

\section{Discussion and final remarks}
\label{sec:final}

The combined impact of positive and negative scale-dependent feedbacks of each biological process is in the core of pattern formation theory~\cite{rietkerk2008regular,cross1993pattern}.
Recently, the idea that the functional form and range of the interactions have a critical role on the spatial organization of organisms has been highlighted~\cite{Dornelas2021,PhysRevLett.98.258101}.  
In this paper, we have connected these  two fundamental results, exploring the consequences that the shape of the interactions  has on the form in which the effects of the different processes superpose and control pattern formation. %
Our results show the existence of a scale-dependent interference that can occur when the kernels, which represent the spatially-extended way of action of each process, are sufficiently compact.

 In Sec.~\ref{sec:linear}, we investigated    rectangular kernels that correspond to interactions restricted to a finite size neighborhood. As predicted  by the linear stability analysis, we observed that nonlocal competition  plays a crucial  role  in pattern formation, determining the wavelength of the long-term patterns ($\sim 2\pi/c$). 
Diffusion and growth, however, can qualitatively interfere in pattern formation. This is possible for  kernels that exhibit a Fourier representation with oscillations. The occurrence and amplitude of these oscillations is associated to the kernel compactness, and their wavelength to kernel range. The mode growth rate that results from the linear stability analysis is a superposition of these oscillatory components that can interfere either constructively or destructively, according to  Eq.~(\ref{eq:dispersion}).
Therefore, diffusion, which is commonly thought to exclusively smooth the concentration fields, can actually promote or forbid pattern formation, depending on the shape and range of the dispersal kernel.
 
Assuming a nonlocal but linear dependence on the  neighborhood, our results initially indicated that only  dispersal could qualitative affect pattern formation through the interference mechanism. 
 Nevertheless,  we remarked that if the actual dynamics of the population in question included additional factors (neglected in our version of the model given by  Eq.~(\ref{eq:fkpp_lin})), then, the  impact of growth could be amplified. This was shown by proposing a more general nonlinear extension, Eq.~(\ref{eq:fkpp_nonl}), including density-dependent feedbacks.
For this case, although competition is still the main driver of pattern formation, both dispersion and facilitation interferences  produce  qualitative changes.
 Furthermore, nonlinearities  allow the facilitation range to impact on pattern  wavelengths, departing from the one  imposed by the competition kernel,  as illustrated in 
Fig~\ref{fig:wavelength}.  

 These results, obtained throughout Secs.~\ref{sec:linear} and \ref{sec:nonlinear}, assumed uniform interactions, represented by a rectangular kernel with variable width. This shape was chosen due to its minimal form and closed analytical representation in Fourier space. 
Our conclusions, however, can be extended to other kernels whose Fourier representation has oscillations that change sign.  As discussed in Sec.~\ref{sec:stability}, the existence of these oscillations in Fourier space depends on   how much the nonlocal interactions are spatially compact. 
For instance, in the case of the so called $q$-exponential form, $f(x)=(1-(1-q)|x|/\ell)^{1/(1-q)}$ (which is zero if $|x|>\ell/(1-q)$ when $q<1$), oscillations appear exactly if $q<1$, in which case $f$ becomes compact-supported (e.g., for the triangular case, $q=0$, $\tilde{f}(k) = [\sin(k\ell)/k\ell]^2$). But note that, for pattern formation, it is crucial that $\mathcal{C}$ (competition kernel)  can assume negative values (to produce a positive peak in $\lambda$), which is only possible for  $q<0$~\cite{Dornelas2021}. 
For the class of stretched-exponential functions, $f(x)\propto e^{-|x|^\alpha}$, the two features (oscillations and non-definite sign of $\tilde{f}$) occur for $\alpha>2$.

 A natural continuation of this research is to investigate the effects of stochastic fluctuations, 
which similarly to nonlinear couplings, can reveal hidden oscillations~\cite{da2014effect}. 
As a further extension, it would be interesting to look for the observed phenomenon in other systems, 
for instance in the dynamics of specific populations  with their particularities~\cite{borgogno2009mathematical,martinez2013vegetation} or even in other reaction-diffusion systems that can contain several nonlocal terms.

 {\bf Acknowledgments:}  
Conselho Nacional de Desenvolvimento Cient\'{\i}fico e Tecnol\'ogico (CNPq) is acknowledged for partial financial support by CA (process 311435/2020-3) and GGP. CA and GGP also acknowledge Coordena\c{c}\~ao de Aperfei\c{c}oamento de Pessoal de Nível Superior (CAPES) (finance code 001).


\begin{thebibliography}{32}
\expandafter\ifx\csname natexlab\endcsname\relax\def\natexlab#1{#1}\fi
\expandafter\ifx\csname bibnamefont\endcsname\relax
  \def\bibnamefont#1{#1}\fi
\expandafter\ifx\csname bibfnamefont\endcsname\relax
  \def\bibfnamefont#1{#1}\fi
\expandafter\ifx\csname citenamefont\endcsname\relax
  \def\citenamefont#1{#1}\fi
\expandafter\ifx\csname url\endcsname\relax
  \def\url#1{\texttt{#1}}\fi
\expandafter\ifx\csname urlprefix\endcsname\relax\def\urlprefix{URL }\fi
\providecommand{\bibinfo}[2]{#2}
\providecommand{\eprint}[2][]{\url{#2}}

\bibitem[{\citenamefont{Tilman et~al.}(1997)\citenamefont{Tilman, Kareiva
  et~al.}}]{tilman1997spatial}
\bibinfo{author}{\bibfnamefont{D.}~\bibnamefont{Tilman}},
  \bibinfo{author}{\bibfnamefont{P.~M.} \bibnamefont{Kareiva}},
  \bibnamefont{et~al.}, \emph{\bibinfo{title}{Spatial ecology: the role of
  space in population dynamics and interspecific interactions}}
  (\bibinfo{publisher}{Princeton University Press}, \bibinfo{year}{1997}).

\bibitem[{\citenamefont{Hanski et~al.}(1999)}]{hanski1999metapopulation}
\bibinfo{author}{\bibfnamefont{I.}~\bibnamefont{Hanski}} \bibnamefont{et~al.},
  \emph{\bibinfo{title}{Metapopulation ecology}} (\bibinfo{publisher}{Oxford
  University Press}, \bibinfo{year}{1999}).

\bibitem[{\citenamefont{Camazine et~al.}(2020)\citenamefont{Camazine,
  Deneubourg, Franks, Sneyd, Theraula, and Bonabeau}}]{camazine2020self}
\bibinfo{author}{\bibfnamefont{S.}~\bibnamefont{Camazine}},
  \bibinfo{author}{\bibfnamefont{J.-L.} \bibnamefont{Deneubourg}},
  \bibinfo{author}{\bibfnamefont{N.~R.} \bibnamefont{Franks}},
  \bibinfo{author}{\bibfnamefont{J.}~\bibnamefont{Sneyd}},
  \bibinfo{author}{\bibfnamefont{G.}~\bibnamefont{Theraula}}, \bibnamefont{and}
  \bibinfo{author}{\bibfnamefont{E.}~\bibnamefont{Bonabeau}},
  \emph{\bibinfo{title}{Self-organization in biological systems}}
  (\bibinfo{publisher}{Princeton University Press}, \bibinfo{year}{2020}).

\bibitem[{\citenamefont{Cross and Hohenberg}(1993)}]{cross1993pattern}
\bibinfo{author}{\bibfnamefont{M.~C.} \bibnamefont{Cross}} \bibnamefont{and}
  \bibinfo{author}{\bibfnamefont{P.~C.} \bibnamefont{Hohenberg}},
  \bibinfo{journal}{Reviews of Modern Physics} \textbf{\bibinfo{volume}{65}},
  \bibinfo{pages}{851} (\bibinfo{year}{1993}).

\bibitem[{\citenamefont{Bonachela et~al.}(2015)\citenamefont{Bonachela,
  Pringle, Sheffer, Coverdale, Guyton, Caylor, Levin, and
  Tarnita}}]{bonachela2015termite}
\bibinfo{author}{\bibfnamefont{J.~A.} \bibnamefont{Bonachela}},
  \bibinfo{author}{\bibfnamefont{R.~M.} \bibnamefont{Pringle}},
  \bibinfo{author}{\bibfnamefont{E.}~\bibnamefont{Sheffer}},
  \bibinfo{author}{\bibfnamefont{T.~C.} \bibnamefont{Coverdale}},
  \bibinfo{author}{\bibfnamefont{J.~A.} \bibnamefont{Guyton}},
  \bibinfo{author}{\bibfnamefont{K.~K.} \bibnamefont{Caylor}},
  \bibinfo{author}{\bibfnamefont{S.~A.} \bibnamefont{Levin}}, \bibnamefont{and}
  \bibinfo{author}{\bibfnamefont{C.~E.} \bibnamefont{Tarnita}},
  \bibinfo{journal}{Science} \textbf{\bibinfo{volume}{347}},
  \bibinfo{pages}{651} (\bibinfo{year}{2015}).

\bibitem[{\citenamefont{Zhao et~al.}(2021)\citenamefont{Zhao, Zhang, Siteur,
  Li, Liu, and van~de Koppel}}]{Zhaoeabe1100}
\bibinfo{author}{\bibfnamefont{L.-X.} \bibnamefont{Zhao}},
  \bibinfo{author}{\bibfnamefont{K.}~\bibnamefont{Zhang}},
  \bibinfo{author}{\bibfnamefont{K.}~\bibnamefont{Siteur}},
  \bibinfo{author}{\bibfnamefont{X.-Z.} \bibnamefont{Li}},
  \bibinfo{author}{\bibfnamefont{Q.-X.} \bibnamefont{Liu}}, \bibnamefont{and}
  \bibinfo{author}{\bibfnamefont{J.}~\bibnamefont{van~de Koppel}},
  \bibinfo{journal}{Science Advances} \textbf{\bibinfo{volume}{7}}
  (\bibinfo{year}{2021}).

\bibitem[{\citenamefont{Heinsalu et~al.}(2013)\citenamefont{Heinsalu,
  Hern\'andez-Garcia, and L\'opez}}]{PhysRevLett.110.258101}
\bibinfo{author}{\bibfnamefont{E.}~\bibnamefont{Heinsalu}},
  \bibinfo{author}{\bibfnamefont{E.}~\bibnamefont{Hern\'andez-Garcia}},
  \bibnamefont{and} \bibinfo{author}{\bibfnamefont{C.}~\bibnamefont{L\'opez}},
  \bibinfo{journal}{Physical Review Letters} \textbf{\bibinfo{volume}{110}},
  \bibinfo{pages}{258101} (\bibinfo{year}{2013}).

\bibitem[{\citenamefont{Maciel and Martinez-Garcia}(2021)}]{Maciel21}
\bibinfo{author}{\bibfnamefont{G.~A.} \bibnamefont{Maciel}} \bibnamefont{and}
  \bibinfo{author}{\bibfnamefont{R.}~\bibnamefont{Martinez-Garcia}},
  \bibinfo{journal}{Journal of Theoretical Biology}
  \textbf{\bibinfo{volume}{530}}, \bibinfo{pages}{110872}
  (\bibinfo{year}{2021}).

\bibitem[{\citenamefont{Murray}(2001)}]{murray2001mathematical}
\bibinfo{author}{\bibfnamefont{J.~D.} \bibnamefont{Murray}},
  \emph{\bibinfo{title}{Mathematical biology II: spatial models and biomedical
  applications}}, vol.~\bibinfo{volume}{3} (\bibinfo{publisher}{Springer New
  York}, \bibinfo{year}{2001}).

\bibitem[{\citenamefont{Rietkerk and Van~de
  Koppel}(2008)}]{rietkerk2008regular}
\bibinfo{author}{\bibfnamefont{M.}~\bibnamefont{Rietkerk}} \bibnamefont{and}
  \bibinfo{author}{\bibfnamefont{J.}~\bibnamefont{Van~de Koppel}},
  \bibinfo{journal}{Trends in Ecology \& Evolution}
  \textbf{\bibinfo{volume}{23}}, \bibinfo{pages}{169} (\bibinfo{year}{2008}).

\bibitem[{\citenamefont{HilleRisLambers
  et~al.}(2001)\citenamefont{HilleRisLambers, Rietkerk, van~den Bosch, Prins,
  and de~Kroon}}]{hillerislambers2001vegetation}
\bibinfo{author}{\bibfnamefont{R.}~\bibnamefont{HilleRisLambers}},
  \bibinfo{author}{\bibfnamefont{M.}~\bibnamefont{Rietkerk}},
  \bibinfo{author}{\bibfnamefont{F.}~\bibnamefont{van~den Bosch}},
  \bibinfo{author}{\bibfnamefont{H.~H.} \bibnamefont{Prins}}, \bibnamefont{and}
  \bibinfo{author}{\bibfnamefont{H.}~\bibnamefont{de~Kroon}},
  \bibinfo{journal}{Ecology} \textbf{\bibinfo{volume}{82}}, \bibinfo{pages}{50}
  (\bibinfo{year}{2001}).

\bibitem[{\citenamefont{Borgogno et~al.}(2009)\citenamefont{Borgogno,
  D'Odorico, Laio, and Ridolfi}}]{borgogno2009mathematical}
\bibinfo{author}{\bibfnamefont{F.}~\bibnamefont{Borgogno}},
  \bibinfo{author}{\bibfnamefont{P.}~\bibnamefont{D'Odorico}},
  \bibinfo{author}{\bibfnamefont{F.}~\bibnamefont{Laio}}, \bibnamefont{and}
  \bibinfo{author}{\bibfnamefont{L.}~\bibnamefont{Ridolfi}},
  \bibinfo{journal}{Reviews of Geophysics} \textbf{\bibinfo{volume}{47}}
  (\bibinfo{year}{2009}).

\bibitem[{\citenamefont{Fisher}(1937)}]{FISHER1937WAVE}
\bibinfo{author}{\bibfnamefont{R.~A.} \bibnamefont{Fisher}},
  \bibinfo{journal}{Annals of Eugenics} \textbf{\bibinfo{volume}{7}},
  \bibinfo{pages}{355} (\bibinfo{year}{1937}).

\bibitem[{\citenamefont{Fuentes et~al.}(2003)\citenamefont{Fuentes, Kuperman,
  and Kenkre}}]{fuentes2003nonlocal}
\bibinfo{author}{\bibfnamefont{M.}~\bibnamefont{Fuentes}},
  \bibinfo{author}{\bibfnamefont{M.}~\bibnamefont{Kuperman}}, \bibnamefont{and}
  \bibinfo{author}{\bibfnamefont{V.}~\bibnamefont{Kenkre}},
  \bibinfo{journal}{Physical Review Letters} \textbf{\bibinfo{volume}{91}},
  \bibinfo{pages}{158104} (\bibinfo{year}{2003}).

\bibitem[{\citenamefont{da~Cunha et~al.}(2011)\citenamefont{da~Cunha, Penna,
  and Oliveira}}]{da2011pattern}
\bibinfo{author}{\bibfnamefont{J.~A.} \bibnamefont{da~Cunha}},
  \bibinfo{author}{\bibfnamefont{A.~L.} \bibnamefont{Penna}}, \bibnamefont{and}
  \bibinfo{author}{\bibfnamefont{F.~A.} \bibnamefont{Oliveira}},
  \bibinfo{journal}{Physical Review E} \textbf{\bibinfo{volume}{83}},
  \bibinfo{pages}{015201} (\bibinfo{year}{2011}).

\bibitem[{\citenamefont{Colombo and Anteneodo}(2012)}]{colombo2012nonlinear}
\bibinfo{author}{\bibfnamefont{E.~H.} \bibnamefont{Colombo}} \bibnamefont{and}
  \bibinfo{author}{\bibfnamefont{C.}~\bibnamefont{Anteneodo}},
  \bibinfo{journal}{Physical Review E} \textbf{\bibinfo{volume}{86}},
  \bibinfo{pages}{036215} (\bibinfo{year}{2012}).

\bibitem[{\citenamefont{Kang et~al.}(2020)\citenamefont{Kang, Ruan, and
  Yu}}]{kang2020age}
\bibinfo{author}{\bibfnamefont{H.}~\bibnamefont{Kang}},
  \bibinfo{author}{\bibfnamefont{S.}~\bibnamefont{Ruan}}, \bibnamefont{and}
  \bibinfo{author}{\bibfnamefont{X.}~\bibnamefont{Yu}},
  \bibinfo{journal}{Journal of Dynamics and Differential Equations}
  (\bibinfo{year}{2020}).

\bibitem[{\citenamefont{Bates}(2006)}]{bates2006some}
\bibinfo{author}{\bibfnamefont{P.~W.} \bibnamefont{Bates}},
  \bibinfo{journal}{Nonlinear Dynamics and Evolution Equations}
  \textbf{\bibinfo{volume}{48}}, \bibinfo{pages}{13} (\bibinfo{year}{2006}).

\bibitem[{\citenamefont{Andreu-Vaillo et~al.}(2010)\citenamefont{Andreu-Vaillo,
  Maz{\'o}n, Rossi, and Toledo-Melero}}]{andreu2010nonlocal}
\bibinfo{author}{\bibfnamefont{F.}~\bibnamefont{Andreu-Vaillo}},
  \bibinfo{author}{\bibfnamefont{J.~M.} \bibnamefont{Maz{\'o}n}},
  \bibinfo{author}{\bibfnamefont{J.~D.} \bibnamefont{Rossi}}, \bibnamefont{and}
  \bibinfo{author}{\bibfnamefont{J.~J.} \bibnamefont{Toledo-Melero}},
  \emph{\bibinfo{title}{Nonlocal diffusion problems}}, \bibinfo{number}{165}
  (\bibinfo{publisher}{American Mathematical Soc.}, \bibinfo{year}{2010}).

\bibitem[{\citenamefont{Turchin}(1998)}]{turchin1998quantitative}
\bibinfo{author}{\bibfnamefont{P.}~\bibnamefont{Turchin}},
  \emph{\bibinfo{title}{Quantitative Analysis of Movement: Measuring and
  Modeling Population Redistribution in Animals and Plants}}, Weimar and Now;
  13 (\bibinfo{publisher}{Sinauer}, \bibinfo{year}{1998}).

\bibitem[{\citenamefont{Mart{\'\i}nez-Garc{\'\i}a
  et~al.}(2013)\citenamefont{Mart{\'\i}nez-Garc{\'\i}a, Calabrese,
  Hern{\'a}ndez-Garc{\'\i}a, and L{\'o}pez}}]{martinez2013vegetation}
\bibinfo{author}{\bibfnamefont{R.}~\bibnamefont{Mart{\'\i}nez-Garc{\'\i}a}},
  \bibinfo{author}{\bibfnamefont{J.~M.} \bibnamefont{Calabrese}},
  \bibinfo{author}{\bibfnamefont{E.}~\bibnamefont{Hern{\'a}ndez-Garc{\'\i}a}},
  \bibnamefont{and}
  \bibinfo{author}{\bibfnamefont{C.}~\bibnamefont{L{\'o}pez}},
  \bibinfo{journal}{Geophysical Research Letters}
  \textbf{\bibinfo{volume}{40}}, \bibinfo{pages}{6143} (\bibinfo{year}{2013}).

\bibitem[{\citenamefont{Dornelas et~al.}(2021)\citenamefont{Dornelas, Colombo,
  L{\'o}pez, Hern{\'a}ndez-Garc{\'i}a, and Anteneodo}}]{Dornelas2021}
\bibinfo{author}{\bibfnamefont{V.}~\bibnamefont{Dornelas}},
  \bibinfo{author}{\bibfnamefont{E.~H.} \bibnamefont{Colombo}},
  \bibinfo{author}{\bibfnamefont{C.}~\bibnamefont{L{\'o}pez}},
  \bibinfo{author}{\bibfnamefont{E.}~\bibnamefont{Hern{\'a}ndez-Garc{\'i}a}},
  \bibnamefont{and}
  \bibinfo{author}{\bibfnamefont{C.}~\bibnamefont{Anteneodo}},
  \bibinfo{journal}{Scientific Reports} \textbf{\bibinfo{volume}{11}},
  \bibinfo{pages}{3470} (\bibinfo{year}{2021}).

\bibitem[{\citenamefont{Pigolotti et~al.}(2007)\citenamefont{Pigolotti,
  L\'opez, and Hern\'andez-Garc\'{\i}a}}]{PhysRevLett.98.258101}
\bibinfo{author}{\bibfnamefont{S.}~\bibnamefont{Pigolotti}},
  \bibinfo{author}{\bibfnamefont{C.}~\bibnamefont{L\'opez}}, \bibnamefont{and}
  \bibinfo{author}{\bibfnamefont{E.}~\bibnamefont{Hern\'andez-Garc\'{\i}a}},
  \bibinfo{journal}{Physical Review Letters} \textbf{\bibinfo{volume}{98}},
  \bibinfo{pages}{258101} (\bibinfo{year}{2007}).

\bibitem[{not()}]{note:numerical}
\bibinfo{note}{Numerical integration was performed using a standard first-order
  Euler scheme with spatial discretization $dx=0.1$, and time step $\delta t =
  10^{-4}$. For the range of parameters used (as specified in figure captions),
  for $t > 100$, the density profiles remain unchanged (relative change smaller
  than $10^{-6}$). In figures, the plotted long-time distribution was extracted
  at $t=400$.}

\bibitem[{\citenamefont{da~Silva et~al.}(2014)\citenamefont{da~Silva, Colombo,
  and Anteneodo}}]{da2014effect}
\bibinfo{author}{\bibfnamefont{L.}~\bibnamefont{da~Silva}},
  \bibinfo{author}{\bibfnamefont{E.}~\bibnamefont{Colombo}}, \bibnamefont{and}
  \bibinfo{author}{\bibfnamefont{C.}~\bibnamefont{Anteneodo}},
  \bibinfo{journal}{Physical Review E} \textbf{\bibinfo{volume}{90}},
  \bibinfo{pages}{012813} (\bibinfo{year}{2014}).

\bibitem[{\citenamefont{Martinez et~al.}(2008)\citenamefont{Martinez,
  Gonz{\'a}lez, and Ter{\c{c}}ariol}}]{martinez2008continuous}
\bibinfo{author}{\bibfnamefont{A.~S.} \bibnamefont{Martinez}},
  \bibinfo{author}{\bibfnamefont{R.~S.} \bibnamefont{Gonz{\'a}lez}},
  \bibnamefont{and} \bibinfo{author}{\bibfnamefont{C.~A.~S.}
  \bibnamefont{Ter{\c{c}}ariol}}, \bibinfo{journal}{Physica A: Statistical
  Mechanics and its Applications} \textbf{\bibinfo{volume}{387}},
  \bibinfo{pages}{5679} (\bibinfo{year}{2008}).

\bibitem[{\citenamefont{Martinez et~al.}(2009)\citenamefont{Martinez,
  Gonz{\'a}lez, and Esp{\'\i}ndola}}]{martinez2009generalized}
\bibinfo{author}{\bibfnamefont{A.~S.} \bibnamefont{Martinez}},
  \bibinfo{author}{\bibfnamefont{R.~S.} \bibnamefont{Gonz{\'a}lez}},
  \bibnamefont{and} \bibinfo{author}{\bibfnamefont{A.~L.}
  \bibnamefont{Esp{\'\i}ndola}}, \bibinfo{journal}{Physica A: Statistical
  Mechanics and its Applications} \textbf{\bibinfo{volume}{388}},
  \bibinfo{pages}{2922} (\bibinfo{year}{2009}).

\bibitem[{\citenamefont{Cabella et~al.}(2011)\citenamefont{Cabella, Martinez,
  and Ribeiro}}]{cabella2011data}
\bibinfo{author}{\bibfnamefont{B.~C.~T.} \bibnamefont{Cabella}},
  \bibinfo{author}{\bibfnamefont{A.~S.} \bibnamefont{Martinez}},
  \bibnamefont{and} \bibinfo{author}{\bibfnamefont{F.}~\bibnamefont{Ribeiro}},
  \bibinfo{journal}{Physical Review E} \textbf{\bibinfo{volume}{83}},
  \bibinfo{pages}{061902} (\bibinfo{year}{2011}).

\bibitem[{\citenamefont{dos Santos et~al.}(2014)\citenamefont{dos Santos,
  Cabella, and Martinez}}]{dos2014generalized}
\bibinfo{author}{\bibfnamefont{L.~S.} \bibnamefont{dos Santos}},
  \bibinfo{author}{\bibfnamefont{B.~C.} \bibnamefont{Cabella}},
  \bibnamefont{and} \bibinfo{author}{\bibfnamefont{A.~S.}
  \bibnamefont{Martinez}}, \bibinfo{journal}{Theory in Biosciences}
  \textbf{\bibinfo{volume}{133}}, \bibinfo{pages}{117} (\bibinfo{year}{2014}).

\bibitem[{\citenamefont{dos Santos et~al.}(2015)\citenamefont{dos Santos,
  Ribeiro, and Martinez}}]{dos2015models}
\bibinfo{author}{\bibfnamefont{R.~V.} \bibnamefont{dos Santos}},
  \bibinfo{author}{\bibfnamefont{F.~L.} \bibnamefont{Ribeiro}},
  \bibnamefont{and} \bibinfo{author}{\bibfnamefont{A.~S.}
  \bibnamefont{Martinez}}, \bibinfo{journal}{Journal of Theoretical Biology}
  \textbf{\bibinfo{volume}{385}}, \bibinfo{pages}{143} (\bibinfo{year}{2015}).

\bibitem[{\citenamefont{Colombo and Anteneodo}(2018)}]{colombo2018nonlinear}
\bibinfo{author}{\bibfnamefont{E.}~\bibnamefont{Colombo}} \bibnamefont{and}
  \bibinfo{author}{\bibfnamefont{C.}~\bibnamefont{Anteneodo}},
  \bibinfo{journal}{Journal of Theoretical Biology}
  \textbf{\bibinfo{volume}{446}}, \bibinfo{pages}{11} (\bibinfo{year}{2018}).

\bibitem[{\citenamefont{Kath}(1984)}]{kath1984waiting}
\bibinfo{author}{\bibfnamefont{W.~L.} \bibnamefont{Kath}},
  \bibinfo{journal}{Physica D: Nonlinear Phenomena}
  \textbf{\bibinfo{volume}{12}}, \bibinfo{pages}{375} (\bibinfo{year}{1984}).

\end{thebibliography}
\end{document}